\documentstyle[aps,preprint]{revtex}
\begin{document}
\draft
\title{Angular position of nodes in the superconducting gap of YBCO }
\author{H. Aubin, K. Behnia, M. Ribault}
\address{Laboratoire de Physique des Solides (associ\'e au CNRS),\\
Universit\'e Paris-Sud, 91405 Orsay, France }
\author{R. Gagnon and L. Taillefer}
\address{Department of Physics, 3600 University Street, McGill University, Montr\'eal, Qu\'ebec, Canada H3A2T8}
\date{\today }
\maketitle

\begin{abstract}
The thermal conductivity of a YBa$_2$Cu$_3$O$_{6.9}$ detwinned single crystal
has been studied as a function of the relative orientation of the crystal
axes and a magnetic field rotating in the Cu-O planes. Measurements were
carried out at several different temperatures below T$_c$ (0.5 K 
\mbox{$<$}
T%
\mbox{$<$}
 25 K) and at a fixed field of 30 kOe. A four-fold symmetry characteristic of
a superconducting gap with nodes at odd multiples of 45 degrees in 
k-space was resolved. Experiments were performed to exclude a possible macroscopic
origin for such a four-fold symmetry such as  sample shape or anisotropic 
pinning. Our results impose an upper limit of 10\% on the weight of the s-wave component
of the essentially d-wave superconducting order parameter of YBCO.
\end{abstract}

\pacs{74.25.Fy, 74.72.Bk, 72.15.Eb}

Over the last years, there has been accumulating evidence that the pairing
state of high temperature superconductors may be the so-called $d_{x^2-y^2}$
state. Among the wealth of experimental data, one can mention the results from
magnetic penetration depth \cite{Zhang,bonn,hardy}, angular resolved
photoemission \cite{Shen}, and nuclear spin relaxation rate \cite{Martindale}%
, which provide strong evidence for nodes in the superconducting gap of the
cuprates. Furthermore, quantum phase interference experiments\cite
{Wollman} which look directly at the symmetry of the order parameter as a
function of its argument, generally confirm the presence of the nodes and
the sign change of the gap function over the Fermi surface.

Most of these experiments suggest, as a recent review by Annett $et\;al.$ 
\cite{Leggett} concludes, that a $d_{x^2-y^2}$ order parameter is the most
plausible candidate to describe the superconducting state in these systems.
However, in the case of optimally-dopped YBa$_2$Cu$_3$O$_{7-\delta }$ (YBCO), it
has been argued that a possible slight s-wave component could arise from the
orthorhombic distortion\cite{monod2}. The in-plane anisotropy observed below
T$_c$ in thermal conductivity\cite{yu} and penetration depth\cite{bonn,hardy}
suggest that the CuO chains play a role in the thermodynamic and
transport properties of the superconducting state. Furthermore, it was
recently pointed out that the in-plane penetration depth anisotropy is
difficult to understand within a proximity model with only a pure $d$-wave
pairing in the Cu-O plane\cite{Carbotte}. One consequence of the existence
of a s-wave component would be a shift of the nodal positions from the
principal diagonal directions over the Fermi surface. Thus, the precise
angular location of the gap nodes on the Fermi surface would be a useful
information in the current debate on the symmetry of the order parameter in
YBCO. 

We present here a study of the thermal conductivity of a YBCO single-crystal, $\kappa$, 
rotating in a magnetic field parallel to the Cu-O planes. A pioneer work by Salamon 
and co-workers\cite{salamon,yu2} showed that thanks to the Andreev scattering 
of quasiparticles by rotating vortices, such an experiment can be a relevant probe
of the k-space anisotropy of the superconducting gap. This work was criticized
by Klemm {\it et al.}\cite{klemm} who contested a microscopic interpretation of the data.
Moreover, error bars on the angular dependence of the longitudinal thermal conductivity
were comparable to the magnitude of the periodical oscillations\cite{salamon}, making it 
difficult, if not impossible, to resolve a four-fold symmetry of $\kappa(\theta)$ which 
is the qualitative distinction of d-wave superconductivity. In this letter, we begin by 
reporting on high-resolution experimental results which establish such a four-fold 
symmetry and then we address objections to a microscopic interpretation of this angular 
dependence. Finally, we show that our data impose an upper limit on the relative 
weight of the s-wave component of the essentially d-wave state of YBCO.

Our  YBCO single crystal was prepared by a self-flux method described elsewhere%
\cite{taillefer}. Measurements were performed in a dilution refrigerator
using a one heater-two-thermometer steady state method. The sample was
anchored at one end to a Cu heat sink, and a heater resistor was attached to
the other. The heat current (J) was injected along the b-axis direction
(i.e. parallel to the chains), and the thermal gradient was measured along
the same direction with two Lakeshore Cernox thermometers. The experimental
setup was rotated at the center of a superconducting coil with the rotation
axis along the c-axis and the magnetic field parallel to the Cu-O planes.
The precise relative orientation of the magnetic field and the crystal was
determined through a Hall probe rigidly fixed on the setup. The maximal
available rotation was about 200 degrees. Below 5 K, due to the slightly
anisotropic magnetoresistance of the thermometers, we used a dynamic method
which allowed us to calibrate the thermometers for each measured $\theta $-direction. 

Fig. 1 shows the thermal conductivity of the YBCO sample as a function of
temperature over a wide temperature range.  A few years ago, it was suggested 
that the steep increase of the thermal conductivity at the superconducting 
transition is induced by the electronic contribution in the Cu-O planes which,
due to the strongly suppressed quasiparticle scattering rate in the superconducting
state, increases rapidly below T$_c$ \cite{yu}. Early evidence for such a 
suppression came from microwave conductivity data\cite{Zhang2}. Later, thermal
Hall effect measurements\cite{krishana} confirmed this pattern and it is now 
generally accepted\cite{hirschfeld} that the peak of the thermal 
conductivity below T$_c$ is in large part due to electrons. In our sample this 
peak is higher ($\frac{\kappa_{max}} {\kappa(T_c)}$ = 2.2) and occurs at 
a lower temperature ($\sim 23$ K) than what has been previously reported\cite{yu}. 
Both these features point to a long maximum quasi-particle mean-free path in 
the sample studied in this work. Because of this large electronic contribution, 
thermal conductivity is a useful probe of the quasi-particle excitation spectrum.

The angular variation of the thermal conductivity of YBCO, $\kappa \left(
\theta \right) $ with $\theta =({\bf b},{\bf H})$, at 24 K is shown in fig.
2. The full range of angular variation ( 0 %
\mbox{$<$}
$\theta $%
\mbox{$<$}
 360 degrees ) was obtained by inverting the direction of the magnetic field.
Thermal conductivity shows a clear fourfold variation superposed on a
hysteretic background. This hysteresis, observed between successive
rotations, is related to the pinning of vortices, which might induce
some variation of the vortex density between two consecutive crossings of the
magnetic field at a $\theta $-direction. This hysteresis continues to
be present at lower temperatures where the pinning is  stronger. But, somewhat
surprisingly, the monotonic background disappears. As shown in fig. 3a, 
the angular variation has a clear fourfold symmetry at 6.8 K. This is the first 
time that such a symmetry of thermal conductivity is unambigously established 
well beyond the experimental resolution in a high-T$_c$ superconductor. 

Next, we consider possible objections to a microscopic interpretation of this
fourfold symmetry such as raised by Klemm {\it et al.}\cite{klemm}. These authors 
suggested that the angular structure might  be ''due to the demagnetization and flux 
pinning effects associated with a rectangular sample''. To check 
this hypothesis, we performed complementary verifications. First, we used an 
alternative experimental procedure by rotating the crystal in the normal phase: 
After each rotation of about 1 degree, the sample was heated to a temperature 
above T$_c$ (93 K) before cooling down to
the measurement temperature. The angular variation obtained in this way
( fig.2) presented the same behavior without the sloping
background. Second, to rule out the possibility of a morphological origin 
-related to the demagnetization for example-, we studied carefully 
the thermal conductivity of a niobium (Nb) sample of similar shape (flat 
rectangular slab) and size and our results reproduce previous  
measurements on Nb\cite{soussa}. As shown
in fig. 3b, the angular variation of the thermal conductivity for this
sample has an obvious twofold symmetry. This anisotropy changes sign with
decreasing temperature reflecting the nature of the dominant quasi-particle
carriers\cite{maki}, but the symmetry always remains twofold. Thus, the
relative orientation of vortices and the applied current, and not the sample
geometry, governs the angular structure seen in fig. 3b. 

Now, with the elimination of a macroscopic origin for the fourfold variation
observed in the case of YBCO, we can turn to the microscopic picture which
was originally formulated by Yu {\it et al. }\cite{yu2} and is based on the
anisotropic scattering of the quasi-particles by the vortices. It is 
well-known\cite{houghton} that the vortices have dramatic effects on the 
heat transport in type II superconductors. Even in the case of the cuprates with a magnetic
field parallel to the basal plane, the coreless vortices lying between the
Cu-O planes can act as scatterers of the BCS-like quasiparticles. This
scattering mechanism was expressed \cite{yu}{\it \ }in terms of Andreev
reflection \cite{andreev} which is a process of retroreflection of
excitations where spatial variations of the amplitude or the phase of the
order parameter induce branch conversion of electron-like excitations into
hole-like excitations, and vice versa. In other words, the excited states
distributed above the superconducting gap exhibit in the presence of a phase
gradient (superfluid flow, v$_s$), a Doppler shift of their energies, E=E$_0-
${\bf p.v}$_s$, so, when a quasiparticle approaches the vortex core, its
energy in the superfluid frame reaches the energy gap and is converted to a
quasihole, halting its contribution to the heat transport. This scattering
process is strongly dependent on the relative orientation of the
quasiparticle momentum ({\bf p}) and the superfluid velocity ({\bf v}$_s$).
The latter is imposed by the magnetic field. Indeed, no Andreev reflection
occurs when the quasiparticle momentum is normal to the supercurrent and
parallel to the magnetic field. Thus, thanks to the directionality of this
scattering mechanism, thermal conductivity can identify the preferential
momentum orientation of the excitations above the superconducting gap.

Qualitatively, the observed behavior, i.e. a fourfold variation with maxima for odd
multiples of 45$^{\circ }$, indicates that there are maxima in the
angular distribution of the quasi-particle momentum in the vicinity of 
\mbox{$\vert$}
k$_{\text{x}}$%
\mbox{$\vert$}
=%
\mbox{$\vert$}
k$_{\text{y}}$%
\mbox{$\vert$}
. When the field is aligned along these particular directions, most quasiparticles 
are not Andreev reflected, and the thermal resistance is lower.

The predominance of the fourfold variation at 6.8 K enables us to compare the
curve to a 2D version of the usual BRT expression of the electronic thermal 
conductivity\cite{yu2}:  
\[
{\bf \kappa }^{qp}_{b}=\frac 1{2\pi ^2ck_BT^2\hbar ^2}\int d^2p%
\frac{\upsilon_{g b }^{2} E_p^2}{\Gamma ({\bf H},{\bf p})} $sech$^2(\frac{E_p}{2k_bT})
\]
where $\upsilon_{g b}$ is the the b-axis component of the group velocity, c is the c-axis 
lattice parameter and the effective rate of scattering 
\[
\Gamma ({\bf H},{\bf p})=\frac 1{\tau _0}(1+\frac{\tau _0}{\tau _{\upsilon _0}}\exp
\left\{ \frac{-m^2a_\upsilon ^2\left[ E_p-\left| \Delta ({\bf p})\right|
\right] ^2}{p_F^2\hbar ^2\ln (a_\upsilon /\xi _0)\sin ^2\psi ({\bf p})}%
\right\} )
\]
depends on  $a_\upsilon$, the intervortex mean distance, and  on $\psi ({\bf p})$, the 
angle between the magnetic field and the quasi-momentum direction ${\bf p}$. 
$\frac 1{\tau _0}$ and $\frac 1{\tau _{\upsilon _0}}$ are respectively zero-field and
maximum Andreev scattering rates. The gap function is a mixture of 
d-wave and s-wave components: 
\[
\Delta =\Delta _0\left\{ r+\frac{\cos (p_xa/\hbar )-\cos (p_ya/\hbar )}{%
1-\cos (p_Fa/\hbar )}\right\} 
\]
where r represents the relative weight of the s component. As mentioned above,
 a finite r would shift the nodal positions from the main diagonal directions
 and the angular interval between the peak positions of the thermal conductivity
 would differ from 90$^{\circ }$. Numerical calculations using realistic physical 
 parameters ($E_F =1 eV$, $\tau _0 =3 \tau_{\upsilon _0} =10^{-12}s$ and 
$\frac {\kappa^{qp}} {\kappa} =0.05$) are shown in fig. 3a for pure $d-wave$ $%
(r=0)$, $d+10\%s(r=0.1)$ and $d+30\%s(r=0.3)$ cases. The interval ($2\theta $%
) between the two peaks is shifted from 90 degrees (pure $d-wave$) to 96 (10
\%) and 108 (30\%), following the relation $2\theta =Arccos(r)$. However,
in our experimental curve, the two peaks are separated by 90$\pm 6^{\circ }$. 
Thereafter, according to our results, the upper limit to r, i.e. the relative 
weight of the s component of the superconducting order parameter, 
($\Delta =\Delta _d(r+\cos (2\theta ))$), is about 0.1. This is significantly 
lower than what is suggested through a recent examination\cite{monod} of 
the in-plane penetration depth anisotropy\cite{Zhang}. On the other hand, it is compatible with the absence of any detectable difference
in the zero-field thermal conductivity along the a-axis and the b-axis at 
very low temperatures which also suggest an essentially d-wave state\cite{behnia}.

We also studied the temperature dependence of this angular variation down to
0.5K. Fig. 4 shows the curves for 24K, 6.8K and 0.8K plotted together. A
striking change in the structure of $\kappa $($\theta $) is detected between
6.8K and 1.2 K where the fourfold symmetry fades away. The minimum for the
field perpendicular to the thermal current (b-axis) develops significantly
(i.e. $\kappa _{//}$ 
\mbox{$>$}
$\kappa _{\perp }$). We have found that this minimum becomes deeper with
decreasing temperature or for increasing magnetic field. The origin of this
crossover between these two regimes (6.8K and 1.2 K) is not yet well
understood.

However, the analysis above neglects the effect of vortices in the plane
on both phonons and quasiparticles in the chains. The former may provide an 
explanation for the twofold variation developed at low temperatures. 
On the other hand, as a careful study of the in-plane anisotropy of thermal 
conductivity at various temperatures suggests\cite{taillefer2}, the chain 
contribution to the thermal conductivity increases significantly below 50K 
passing by a maximum around 15 K.
In our geometry, with the heat current applied along the chain orientation, a
two-fold angular structure due to the scattering of chain quasi-particles by
vortices could well be present. Additional experiments of this type on a-oriented
YBCO crystals will be necessary to elucidate the effect of a rotating magnetic 
field on different types of heat carriers at different temperatures.

In conclusion, our results support the existence of nodes in the superconducting
gap at angular positions close to what is expected for a purely d-wave state and 
impose an upper limit of about 0.1 to the relative weight of the s component.
 
We are grateful to G. Bellessa for providing us the niobium crystal.
This Work was funded in part by NSERC of Canada, FCAR of Qu\'ebec and
the Canadian Institute for Advanced Research. L.T.
acknowledges the support of the Alfred P. Sloan Foundation.

\begin{figure}[tbp]
\caption{ Main Panel: The temperature depencence of the thermal conductivity
in a log/log plot. Insert shows a linear plot. Note the sharp increase at T$_c
$ leading to a peak at 23 K.}
\label{fig1}
\end{figure}

\begin{figure}[tbp]
\caption{Open squares and circles show the angular variation of the thermal conductivity
at T=24 K and H= 30 kOe. There are minima for the field parallel and perpendicular
to the b-axis direction (heat current), and maxima when the field is parallel to 
the main diagonal directions. Arrows indicate the rotating direction. Solid circles
indicate the results obtained with an alternative procedure of field-cooling at 
every angle (see text).} 
\label{fig2}
\end{figure}

\begin{figure}[tbp]
\caption{The fourfold variation of the thermal conductivity in
YBCO (a) is compared with the twofold variation for the Nb crystal (b). For
YBCO the numerical results are shown for a pure d-wave, $d+10\%s-wave(r=0.1)$
(solid line) and $d+30\% s-wave(r=0.3)$ (dashed line). For Nb the solid line
is a fit using the  cosine-square dependence expected for a conventional gap.}

\label{fig3}
\end{figure}

\begin{figure}[tbp]
\caption{Angular variation of the thermal conductivity at 24 K, 6.8 K and 0.8 K,
at H= 30 kOe.}
\label{fig4}
\end{figure}


\begin{references}
\bibitem{Zhang}  Kuan Zhang {\it et al.}, Phys. Rev. Lett. {\bf 73}, 2484
(1994).

\bibitem{bonn}  D. A. Bonn {\it et al.}, Phys. Rev. Lett. {\bf 68}, 2390 (1992).

\bibitem{hardy}  W. N. Hardy {\it et al.}, Phys. Rev. Lett. {\bf 70}, 3999 (1993).

\bibitem{Shen} Z.-X. Shen {\it et al.}, Phys. Rev. Lett. {\bf 70}, 1553 (1993);
H. Ding {\it et al.}, Phys. Rev. Lett. {\bf 74}, 2784 (1995).

\bibitem{Martindale}  J.A. Martindale {\it et al.}, Phys. Rev. Lett. {\bf 68},
702 (1992).

\bibitem{Wollman}  D.A. Wollman {\it et al}., Phys. Rev. Lett. {\bf 71},
2134, (1993); C. C. Tsuei {\it et al}., Phys. Rev. Lett. {\bf 73}, 593, (1994).

\bibitem{Leggett}  James Annett, Nigel Goldenfeld and Anthony J. Leggett
[http://xxx.lanl.gov/abs/cond-mat/9601060], to appear in Physical Properties
of High Temperature Superconductors, Vol. 5, D.M.\ Ginsberg (ed.), (World
Scientific, Singapore , 1996).

\bibitem{monod2}  See for example M.T. B\'{e}al-Monod and K. Maki, Phys.
Rev. B, {\bf 53}, 5775 (1996) and references therein.

\bibitem{yu}  R. C. Yu {\it et al.}, Phys. Rev. Lett. {\bf 71}, 1657 (1992).

\bibitem{Carbotte}  W. Atkinson, J.P. Carbotte and C. O'Donovan
[http://xxx.lanl.gov/abs/cond-mat/cond-mat/9604104], To appear in the
proceeding of the BICAS Summer School on ''Symmetry of the Order Parameter
in High-Temperature Superconductors.''

\bibitem{salamon} M.B. Salamon {\it et al.}, J. of Superconductivity, 
{\bf 8}, 449, (1995).

\bibitem{yu2}  F. Yu {\it et al.}, Phys. Rev. Lett. {\bf 74}, 5136 (1995); 
 {\bf 77}, 3059 (1996).

\bibitem{klemm}  R. A. Klemm {\it et al.}, Phys. Rev. Lett. {\bf77}, 3056 (1996).

\bibitem{taillefer}  R. Gagnon, C. Lupien and L. Taillefer, Phys. Rev. B 
{\bf 50}, 3458 (1994).

\bibitem{Zhang2}  Kuan Zhang {\it et al.}, Phys. Rev. Lett. {\bf 73}, 2484
(1994).

\bibitem{krishana}  K. Krishana, J.M. Harris and N.P. Ong, Phys. Rev. Lett.
{\bf 75}, 3529 (1995).

\bibitem{hirschfeld} P.J. Hirschfeld and W.O. Putikka, Phys. Rev. Lett. 
{\bf 77}, 3909 (1996).

\bibitem{soussa}  J. Lowell and J.B. Sousa, J. Low Temp. Physics {\bf3}, 65 (1970).

\bibitem{maki}  Kazumi Maki, Phys. Rev. {\bf 158}, 397 (1967).

\bibitem{houghton}  A. Houghton and K. Maki, Phys. Rev. B {\bf 4}, 843 (1971);
 Robert M. Cleary, Phys. Rev. {\bf 175}, 587, (1968).

\bibitem{andreev}  A.F. Andreev, Sov. Phys. JETP, {\bf 19}, 1228 (1964).

\bibitem{monod}  M.T. B\'{e}al-Monod and K. Maki, to be published in Phys.
Rev. B.

\bibitem{behnia}  K. Behnia {\it et al.}, Synth. Metals {\bf 71}, 1611 (1995).

\bibitem{taillefer2}  Robert Gagnon, Song Pu, Brett Ellman, and Louis Taillefer,
 to be published.

\end{references}
\end{document}